\documentclass[journal]{IEEEtran}[10pt]
\usepackage{blindtext, graphicx}
\usepackage{graphicx}
\usepackage{amsmath}
\usepackage{relsize}
\usepackage{amssymb}
\usepackage{tikz}
\usepackage{float}
\usepackage{mathtools}
\usepackage{algorithm}
\usepackage[noend]{algpseudocode}
\usetikzlibrary{positioning}
\usepackage{caption}
\usepackage{subfig}
\usepackage{gensymb}

\usepackage{multirow}
\usepackage{tablefootnote}
\usepackage{fixltx2e}

\makeatletter
\def\ps@IEEEtitlepagestyle{
  \def\@evenfoot{}
}
\newcommand{\superscript}[1]{\ensuremath{^{\textrm{#1}}}}

\begin{document}

\pagestyle{empty}
\thispagestyle{empty}

\title{A Highly Accelerated Parallel Multi-GPU based Reconstruction Algorithm  for Generating Accurate Relative Stopping Powers}
\author{
Paniz~Karbasi\superscript{[1]},
Ritchie~Cai\superscript{[1]},
Blake~Schultze\superscript{[1]},
Hanh~Nguyen\superscript{[1]},
Jones~Reed\superscript{[1]},
Patrick~Hall\superscript{[1]},
Valentina~Giacometti\superscript{[2],[6]},
Vladimir~Bashkirov\superscript{[6]},
Robert~Johnson\superscript{[3]},
Nick~Karonis\superscript{[4]}
Jeffrey~Olafsen\superscript{[5]},
Caesar~Ordonez\superscript{[4]}
Keith~E.~Schubert\superscript{[1],[6]}~\IEEEmembership{Senior Member,~IEEE},
Reinhard~W.~Schulte\superscript{[6]}~\IEEEmembership{Member,~IEEE}
\thanks{The research in proton CT is supported by the National Institute of Biomedical Imaging and Bioengineering (NIBIB), and the National Science Foundation (NSF), award Number R01EB013118, and the United States - Israel Binational Science Foundation (BSF) grant nos. 2009012 and 2013003. The content of this paper is solely the responsibility of the authors and does not necessarily represent the official views of the National Institute of Biomedical Imaging and Bioengineering or the National Institutes of Health.}
\thanks{The support of UT Southwestern and State of Texas through a Seed Grants in Particle Therapy award is gratefully acknowledged.}
\thanks{We gratefully acknowledge Brian Sitton from Baylor University for providing technical support for the use of Kodiak Cluster at Baylor. Also, we express our special thanks to the IBM Poughkeepsie technical staff for supporting the use of Power8 HPC Shared Cluster.}
\thanks{[1] Paniz Karbasi, Ritchie Cai, Blake Schultze, Hanh Nguyen, Jones Reed, Patrick Hall and Keith Schubert are with the Department of Electrical and Computer Engineering, Baylor University, Waco, TX 76798 USA, email: Paniz\_Karbasi@baylor.edu, Ritchie\_Cai@baylor.edu, Blake\_Schultze@baylor.edu, Hanh\_Nguyen@baylor.edu, Jones\_Reed@baylor.edu, Patrick\_Hall@baylor.edu,  Keith\_Schubert@baylor.edu}%
\thanks{[2] Valentina Giacometti is with the Center for Medical Radiation Physics, University of Wollongong, Wollongong, NSW, Australia, email: valentina8giacometti@gmail.com}%
\thanks{[3] Robert Johnson is with the Santa Cruz Institute for Particle Physics, University of California, Santa Cruz, Santa Cruz, CA 95064, USA, email:  rjohnson@ucsc.edu}%
\thanks{[4] Nick Karonis and Caesar Ordonez  are with the Department of Computer Science, Northern Illinois University, DeKalb, IL 60115 email: karonis@niu.edu,  cordonez@cs.niu.edu}%
\thanks{[5] Jeffrey Olafsen is with the Department of Physics, Baylor University, Waco, TX 76798 USA, email:
Jeffrey\_Olafsen@baylor.edu}%
\thanks{[6] Valentina Giacometti, Vladimir Bashkirov (vabashkirov@llu.edu), Keith Schubert, and Reinhard Schulte (rschulte@llu.edu), are with Loma Linda University, Loma Linda, CA 92350}

}
%
%


\maketitle

\section{Introduction}
Proton Computed Tomography (pCT) is a growing imaging technology  in proton therapy planning. By addressing the range uncertainty problem, pCT images suggest more accurate treatment plans than X-ray CT images~\cite{BSCEWSPRMS08}. The pCT collaboration has developed a proton CT scanner including a silicon-based tracking system and a multi-stage scintillating energy detector for measuring the water equivalent path length (WEPL) of individual protons~\cite{johnson2017results}. By using low-dose proton emission, calculating the individual proton's most likely path~\cite{SPTS08}, and knowing their energy loss, large and sparse linear system of equations $Ax = b$ can be written where $A$ is a $m \times n$ matrix containing the path data, $b$ is the $m \times 1$ WEPLs vector, and $x$ is the $n \times 1$ relative stopping powers (RSP) vector. Using the FBP image as the initial iterate, one can iteratively solve the system $Ax = b$ for generating the 3D map of the RSPs  to be used  in the treatment planning. 

Various concepts ranging from advanced detector designs to appropriately selecting the preprocessing techniques and parameters of the reconstruction algorithm have a major impact on the quality and accuracy of the pCT images~\cite{schultze2015reconstructing},~\cite{dedes2017application}. Calculating  accurate RSPs  is  of  great  importance,  but  to  take advantage of the pCT systems in a clinical setup, there is also a great need for  real-time algorithms that can process hundreds  of  millions  of  protons  in  concise  time frames not more than a few minutes.

The pCT problem fits well within the Single Instruction Multiple Threads (SIMT) parallel programming paradigm of  Graphics Processing Units (GPUs) since we treat each proton as an individual thread that can be processed in parallel. Although there is a great need for small pCT runs as a validation technique in clinics, when it comes to imaging an adult torso, because of the greatly increased problem size, the reconstruction time  grows proportional to the four thirds power of the reconstructed object's volume. In order to achieve the time frames that are clinically meaningful (i.e. in under 5 minutes), we have proposed a  reconstruction technique which takes advantage of  systems with at least 2 GPUs and generates the 3D map of highly accurate RSP values within 40 seconds for 116 million and 79 seconds for 261 million proton histories. Moreover, based on the experimental results, our proposed reconstruction algorithm  runs faster than the expected  speedup on the Nvidia K40 GPUs which demonstrates it is a reasonable and economical alternative for the clinical pCT systems.

\section{Related Work}
The rapid evolution of GPUs in the recent years has contributed to design and implementation of accelerated processing algorithms of many real world applications. Real-time reconstruction of the 3D maps of RSP measurements is a key factor that needs to be met in pCT imaging.  In~\cite{duffin2012analysis} it has been demonstrated that the reconstruction time can be reduced from 7 hours on a single machine to 53 seconds using a GPU cluster for a dataset of size $131$ million protons. In~\cite{karonis2013distributed} it has been further demonstrated that image reconstruction in pCT can be accelerated through a hybrid approach that uses both Message Passing Interface (MPI) and GPUs. Using this approach on the same cluster the reconstruction rumtime has been  improved and reduced to 43 seconds for a similar size of dataset. In a recent study in~\cite{ordonez2017real}, it is shown using the same approach as the first evaluation of the pCT software in~\cite{karonis2013distributed}, the execution time for generating accurate RSP values   for a dataset of size $131$ million protons is almost $30$ seconds running on $60$ processors ($60$ CPU cores + $60$ GPUs). The method we have proposed in this paper, removes the cost of having a GPU cluster, and generates  accurate RSPs with only a single computer and two P100 GPUs thus reducing the time to $40$ seconds for a simulated CTP404 dataset of size $116$ million protons.

\begin{figure*}[ht!]
  \centering \includegraphics[width=15cm,height=4cm]{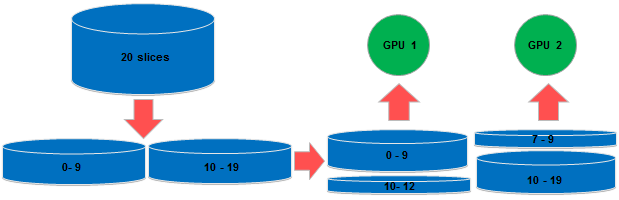}
  \caption{Schematic drawing of Algorithm~\ref{alg:multi-gpu} with two GPUs}
  \label{f:1}
\end{figure*}

\section{Methodology}
In order to benefit from the Nvidia Tesla GPUs, we have developed a fast and easy to implement reconstruction algorithm that can run on systems with at least 2 GPUs with the advantage of eliminating the need of data transfers among the GPUs. The algorithm we have designed relies on the following two facts:

\begin{enumerate}
    \item only a small fraction of protons intersect multiple slices along the vertical axis
    \item within a slice, only a small portion of protons pass through and are used in reconstruction of that specific slice
\end{enumerate}

Based on the first observation which is reported with details in Table~\ref{t:1}, we know that a proton's path does not have a significant deviation along the $z$ axis or vertical slices of the reconstruction volume. This enables us to consider an algorithm without having a significant concern about the GPU-to-GPU data transfers which could potentially add some considerable timing constrains.

The second observation allows us to split the protons among the available GPUs such that each GPU reconstructs a portion of the 3D image. Each GPU only needs some fraction of the protons and this lowers the amount of required memory and processing time per GPU leading to a faster reconstruction time overall.

\begin{table}[b]
\renewcommand{\arraystretch}{1.3}
\caption{Percentage of protons passing through image slices along the $z$ axis for the simulated and experimental CTP404 phantom with the $0.25$ centimeters slice thickness} 
  \label{t:1}
  \centering
  \begin{tabular}{| c || c || c |}
  \hline
  \bfseries \# of slices  & \bfseries Simulated CTP404 & \bfseries Experimental CTP404 \\ 
  \hline \hline
   \bfseries 1 & 32 & 42 \\ \hline
 \bfseries 2 & 43  & 36 \\ \hline 
 \bfseries 3 & 19  & 15 \\ \hline 
  \bfseries 4 & 4 & 3\\ \hline
  $\textbf{$\ge$ 5}$ & 2 & 4\\
   \hline  
  \end{tabular}
\end{table}

The general structure of the proposed multi-GPU based reconstruction technique can be seen in Algorithm~\ref{alg:multi-gpu}. The first step is the division of the reconstruction volume into several overlapping regions based on the number of available GPUs, while each region is assigned to a unique key. The overlap is set at twice the slices for a $96\%$  coverage (Table~\ref{t:1}). In the second step, each proton is associated with the image region  entered to and  exited from. In the third step, based on the number of protons that pass through each image region, the  required memory for each proton  is allocated on each GPU. Lastly (fourth step), we perform a standard iterative solver~\cite{karbasi2015incorporating} on each GPU. Finally, in the overlap of two regions, where there are two reconstructions of each slice, we select the reconstruction that is closest to the non-overlapped slices of its region.  For example, in the reconstruction presented in this work, there are two regions, one containing slices $0-12$ and the other containing slices $7-19$.  The overlap corresponds to slices $7-12$, and the final image will be made of slices $0-9$ from the first region and $10-19$ of the second. The high level description of Algorithm~\ref{alg:multi-gpu} is illustrated in Fig.~\ref{f:1}.

\begin{algorithm}[h]
\caption{multi-GPU based reconstruction algorithm}\label{euclid}
\begin{algorithmic}[0]
\Procedure{Multi-GPU Based Reconstruction}{}
\State 1: setupImageRegions()
\State 2: identifyProtonsPassingRegions()
\State 3: allocateMemoryForProtonsPerGPU()
\State 4: iterativelySolvePerGPU()
\State 5: selectSlice()
\EndProcedure
\end{algorithmic}
\label{alg:multi-gpu}
\end{algorithm}

\section{Experiments and Results}
In this section, we compare the runtime of the proposed multi-GPU based reconstruction technique on three different systems using single and double GPUs, and compare the accuracy of the reconstructed RSP values generated by Algorithm~\ref{alg:multi-gpu} and the standard single GPU technique. 

\subsection{Systems and Datasets}
In order to analyze the performance of the proposed  algorithm discussed in the previous section, we have tested the algorithm on three different platforms: 2 Nvidia K40s on a Xeon, Cray's 2 P100 GPUs, and IBM's P100 GPU.

Angle intervals of $4$-degree per projection ($90$ projections for a full rotation scan) and continuous angle increments were used for the reconstruction of simulated and experimental CTP404 phantom datasets, respectively. Also, both of these datasets are composed of $20$ slices in the vertical direction while the slice thickness is 0.25 cm. 

The simulated data were obtained using the Geant4 based software simulation platform described in~\cite{giacometti2017software}. The simulated and experimental CTP404 data are composed of $116$ million and $261$ million proton histories, respectively, later reduced using data cleaning techniques (e.g. identifying and removing statistical outliers). The actual number of protons used by the iterative solver is $21$ and $78$ million protons for the simulated and experimental data respectively. When using the proposed reconstruction algorithm described in Algorithm~\ref{alg:multi-gpu}, there are different number of protons passing through each image region which are reported in Table~\ref{t:2}.

Based on the results in Table~\ref{t:2}, protons are evenly distributed between the two image regions for the simulated data, while the top image region of the experimental data compasses twice the number of protons that pass through the bottom image region. Execution times reported in the next section heavily depend on the number of protons traversing the image regions and the distribution of data. 

\begin{table}[H]
\renewcommand{\arraystretch}{1.3}
\caption{Number of protons (millions) in each image region used by the GPUs to perform the iterative solver} 
  \label{t:2}
  \centering
  \begin{tabular}{| c || c || c |}
  \hline
  \bfseries Data  &  \bfseries Slices 0-12 & \bfseries Slices 7-19   \\ 
  \hline \hline
   \bfseries Simulated CTP404 & 14 & 14 \\ \hline
   \bfseries Experimental CTP404 & 62  & 33\\  
   \hline  
  \end{tabular}
\end{table}

\subsection{Performance and Execution Time}
In order to test the performance of our proposed reconstruction algorithm, we used two different block sizes $3200$ and $320000$ for the experimental data and $320000$ for the simulated data. The block size is the number of protons to be processed in parallel based on the block iterative technique Diagonally-Relaxed Orthogonal Projections, (DROP~\cite{PSCBMSR10}). When it comes to the performance of the iterative solver, in general, larger block sizes decrease the runtime of a single iteration but not necessarily generate accurate RSPs. Here we have reported the timings of both $3200$ and $320000$ for the experimental data for comparison purpose (see Tables~\ref{t:5} and~\ref{t:7}). 

Based on the timings in Table~\ref{t:3}, the performance of Algorithm~\ref{alg:multi-gpu} is about $1.75$ times faster for the simulated  data on  K40 while based on Table~\ref{t:5}, the iterative solver with two GPUs is  $1.3$ times faster than the single-GPU iterative solver for the experimental data. The reason comes from the fact that the number of protons used by the top image region (slices $0-12$) is only $1.25$ times less than the total number of protons used by the single-GPU iterative solver for experimental data (Table~\ref{t:2}) and since  slice number $9$ comes from the top image region with $62$ million protons, there is not a significant performance improvement with the iterative solver using two GPUs on the experimental data. 

An important observation regarding the timings in Table~\ref{t:3} is that on P100 system, the runtime of the iterative solver is about $1.53$ times faster with the proposed method and  is similar to the expected speedup $\frac{21}{14}$ or $1.5$. On the other hand, the speedup of the proposed method on K40 system is about $1.75$, which is greater than the expected one. The reason comes from the difference between the memory sizes of K40 and P100 GPUs. In fact, the smaller memory of the K40 system in comparison to P100, leads to  efficient usage of the cache on K40 leading to a faster runtime of a single iteration. The total runtime of the pCT software including the data reads and data cleaning techniques is reported in Tables~\ref{t:4} and ~\ref{t:6} for the simulated and experimental data respectively.

\begin{table}[H]
\renewcommand{\arraystretch}{1.3}
\caption{Runtime (second) comparison of Algorithm~\ref{alg:multi-gpu} (step 4) with $1$ iteration of the single-GPU iterative solver on simulated data with block size of $320000$}
\label{t:3}
\centering
\begin{tabular}{lcc}
\hline
& \bfseries  Current Algorithm & \bfseries Proposed Algorithm\\
\hline \hline
\bfseries K40 & 13.3  & 7.6 \\
\bfseries P100 & 2.3 (IBM $\&$ Cray) & 1.5 (Cray) \\ \hline
\end{tabular}
\end{table}

\begin{table}[H]
\renewcommand{\arraystretch}{1.3}
\caption{Total reconstruction runtime (second) of simulated data converging after $5$ and $3$ iterations (Table~\ref{t:3}) of the iterative solver for  single-GPU and Algorithm~\ref{alg:multi-gpu}, respectively}
\label{t:4}
\centering
\begin{tabular}{lcc}
\hline
& \bfseries  Current Algorithm & \bfseries Proposed Algorithm\\
\hline \hline
\bfseries K40 & 154  & 96 \\
\bfseries P100 & 45 (IBM $\&$ Cray) & 40 (Cray) \\ \hline
\end{tabular}
\end{table}

\begin{table}[H]
\renewcommand{\arraystretch}{1.3}
\caption{Runtime (second) comparison of Algorithm~\ref{alg:multi-gpu} (step 4) with $1$ iteration of the single-GPU iterative solver on experimental data with block size of $3200$}
\label{t:5}
\centering
\begin{tabular}{lcc}
\hline
& \bfseries  Current Algorithm & \bfseries Proposed Algorithm\\
\hline \hline
\bfseries K40 & 79.7  & 60.8 \\
\bfseries P100 & 10.2 (IBM $\&$ Cray) & 7.1 (Cray) \\ \hline
\end{tabular}
\end{table}

\begin{table}[b]
\renewcommand{\arraystretch}{1.3}
\caption{Total reconstruction runtime (second) of experimental data converging after $5$ and $3$ iterations (Table~\ref{t:5}) of iterative solver for  single-GPU and Algorithm~\ref{alg:multi-gpu}, respectively}
\label{t:6}
\centering
\begin{tabular}{lcc}
\hline
& \bfseries  Current Algorithm & \bfseries Proposed Algorithm\\
\hline \hline
\bfseries K40 & 521  & 311 \\
\bfseries P100 & 106 (IBM $\&$ Cray) & 79 (Cray) \\ \hline
\end{tabular}
\end{table}

\begin{table}[b]
\renewcommand{\arraystretch}{1.3}
\caption{Runtime (second) comparison of Algorithm~\ref{alg:multi-gpu} (step 4) with $1$ iteration of the single-GPU iterative solver on experimental data with block size of $320000$}
\label{t:7}
\centering
\begin{tabular}{lcc}
\hline
& \bfseries  Current Algorithm & \bfseries Proposed Algorithm\\
\hline \hline
\bfseries K40 & 36.8  & 31.2 \\
\bfseries P100 & 4.9 (IBM $\&$ Cray) &  3.6 (Cray) \\ \hline
\end{tabular}
\end{table}

\subsection{Accuracy and Image Quality}
\begin{figure}[H]
\subfloat[Simulated CTP404\label{sfig:testa}]{%
  \includegraphics[height=4.1cm,width=.48\linewidth]{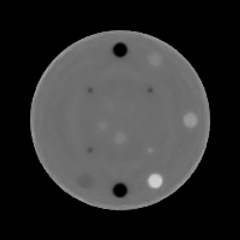}%
}\hfill
\subfloat[Experimental CTP404\label{sfig:testa}]{%
  \includegraphics[height=4.1cm,width=.48\linewidth]{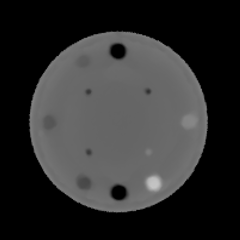}%
}
\caption{Reconstructed images of the CTP404 phantom}
\label{f:2}
\end{figure}

Simulated and experimental reconstructed images are shown in Fig.~\ref{f:2}(a) and~\ref{f:2}(b), respectively. The slices here presented, belong to the top image region (slice number $9$ out of $20$). The mean RSP values of the different inserts in these images are reported in Tables~\ref{t:8} and~\ref{t:9} for the simulated and experimental data, respectively. The results in Tables~\ref{t:8} and~\ref{t:9} show a very good agreement between the RSP values reconstructed using Algorithm~\ref{alg:multi-gpu} with three overlapping slices and the so-called true RPS~\cite{giacometti2017software}.

\begin{table}[H]
\renewcommand{\arraystretch}{1.3}
\caption{Reconstructed RSPs of the simulated CTP404 }
\label{t:8}
\centering
\begin{tabular}{|c||c||c||c|}
\hline
\bfseries Insert &  \bfseries True RSP &   \bfseries $1$ GPU ($\%$ err.)&  \bfseries $2$ GPUs ($\%$ err.) \\ 
\hline\hline
\bfseries PMP   & 0.883 & 0.886 (0.33) & 0.887 (0.45) \\    \hline
\bfseries LDPE   &  0.980 & 0.986 (0.61) & 0.988 (0.82)\\    \hline
\bfseries Polystyrene  & 1.024 & 1.032 (0.78)& 1.033 (0.87)\\    \hline
\bfseries Acrylic   & 1.160 & 1.163 (0.25)& 1.162 (0.17)\\  \hline
\bfseries Delrin   & 1.359 & 1.349 (-0.73)& 1.347 (-0.88)\\  \hline
\bfseries Teflon   & 1.790 & 1.7895 (-0.03) & 1.786 (-0.22)\\ \hline

\end{tabular}
\end{table}

\begin{table}[H]
\renewcommand{\arraystretch}{1.3}
\caption{Reconstructed RSPs of the experimental CTP404 }
\label{t:9}
\centering
\begin{tabular}{|c||c||c||c|}
\hline
\bfseries Insert &  \bfseries True RSP &  \bfseries $1$ GPU ($\%$ err.) &  \bfseries$2$ GPUs ($\%$ err.)  \\
\hline\hline
\bfseries PMP   & 0.883 & 0.894 (1.24) & 0.897 (1.59) \\    \hline
\bfseries LDPE   & 0.980 &  0.989 (0.92)& 0.990 (1.02)\\    \hline
\bfseries Polystyrene  & 1.024 & 1.033 (0.87)& 1.034 (0.97) \\    \hline
\bfseries Acrylic   & 1.160 & 1.171 (0.94) & 1.173 (1.12) \\  \hline
\bfseries Delrin   & 1.359 & 1.345 (-1.03) & 1.343 (-1.17)\\  \hline
\bfseries Teflon   & 1.790 & 1.784 (-0.33) & 1.781 (-0.50)\\ \hline

\end{tabular}
\end{table}

Finally, note that the RSP values converge only after three iterations of the iterative solver with Algorithm~\ref{alg:multi-gpu} running on a double GPU system, while it takes up to five iterations when running on a single GPU system.

\section{Conclusion}
Proton computed tomography is an evolving imaging technique currently under investigation to improve the accuracy of the proton treatment planning. Achieving the clinical timing constraints is one of the key goals for developing pCT softwares, especially when reconstructing large objects such as an adult torso. 

In this paper, we  proposed a fast and easy reconstruction technique that generates the 3D map of  accurate RSP values in a very short amount of time with only two GPUs. 

Based on the experiments, our proposed method surpasses the expected speedup on the Nvidia K40 GPUs which is a significant benefit that suggests an efficient and more economical hardware for a pCT system to be used in a clinical setup. Another important advantage of the proposed method, is that it can be easily be modified to be implemented on more than two GPUs, which we expect to further improve the timing constraints for exceedingly large datasets.


%


\ifCLASSOPTIONcaptionsoff
  \newpage
\fi



%

\bibliographystyle{IEEEtran}
\bibliography{pct}

%




\end{document}